\documentclass[11pt]{article}
\usepackage[margin=1in]{geometry}
\usepackage{amsmath,amssymb,amsthm}
\usepackage{booktabs}
\usepackage{graphicx}
\usepackage{multirow}
\usepackage{array}
\usepackage{algorithm}
\usepackage{algpseudocode}
\usepackage{hyperref}
\usepackage{float}
\usepackage{caption}
\usepackage{subcaption}
\usepackage[authoryear,round]{natbib}
\csname title\endcsname{The impact of multi-agent debate protocols on debate quality: a controlled case study}
\author{Ramtin Zargari Marandi\\\href{https://orcid.org/0000-0001-9233-1656}{ORCID: 0000-0001-9233-1656}}
\date{March 17, 2026}

\begin{document}
\maketitle

\begin{abstract}
In multi-agent debate (MAD) systems, performance gains are often reported; however, because the debate protocol (e.g., number of agents, rounds, and aggregation rule) is typically held fixed while model-related factors vary, it is difficult to disentangle protocol effects from model effects. To isolate these effects, we compare three main protocols, Within-Round (WR; agents see only current-round contributions), Cross-Round (CR; full prior-round context), and novel Rank-Adaptive Cross-Round (RA-CR; dynamically reorders agents and silences one per round via an external judge model), against a No-Interaction baseline (NI; independent responses without peer visibility). In a controlled macroeconomic case study (20 diverse events, five random seeds, matched prompts/decoding), RA-CR achieves faster convergence than CR, WR shows higher peer-referencing, and NI maximizes Argument Diversity (unaffected across the main protocols). These results reveal a trade-off between interaction (peer-referencing rate) and convergence (consensus formation), confirming protocol design matters. When consensus is prioritized, RA-CR outperforms the others.
\end{abstract}


\section{Introduction}
Multi-agent large language model (LLM) systems often use debate-style interaction to improve output quality \citep{du2024debate,liang2024divergent,wu2024autogen,li2023camel,hong2023metagpt,fan2026imad,zhu2026demystifying}. These systems are a form of collaborative reasoning, where sequential debate and communication topology play a central role in shaping outcomes. Recent work on LLM-based agents uses the term \emph{agent} broadly to include role-conditioned LLM instances that interact with other components, maintain local state, or participate in a larger control loop, even when they are not fully autonomous tool-using systems \citep{liu2024dylan,yehudai2025survey}. In this paper, \emph{agent} is used in that operational sense: each agent is a role-assigned LLM instance that produces messages, receives protocol-dependent peer context, and participates in a multi-turn deliberation process. Under that definition, the setting studied here is described as \emph{multi-agent debate} and as a role-playing agents framework.

Although some recent work evaluates specific debate protocol variations, such as sparse communication topologies \citep{li-etal-2024-improving-multi}, protocols are often held fixed or not systematically ablated as a primary experimental factor across the literature. Related work sometimes uses the broader terms \emph{communication protocol} or \emph{communication regime} for closely related design choices \citep{pham2023cipher}. This confounding makes it difficult to cleanly distinguish protocol effects from model-capability effects. A second limitation is the lack of guidance on when increased coordination justifies additional latency and token cost. More generally, a debate protocol determines what peer evidence an agent can see and when it can react to that evidence, thus protocol design can plausibly change both interaction quality and convergence behavior.

The debate protocol is treated as the main variable in this work. Three primary protocols are compared under matched prompts and decoding settings: \textbf{Within-Round} (WR; agents see only contributions from the current round), \textbf{Cross-Round} (CR; agents receive full context from prior rounds), and \textbf{Rank-Adaptive Cross-Round} (RA-CR; cross-round context with quality-biased agent-turn order; see Methods for details). A \textbf{No-Interaction baseline} (NI) is also included, defined as independent agent responses without peer-message visibility; this baseline is used for calibration and context, not as a primary confirmatory target. The key design factors are defined as (i) the amount of context available to each agent and (ii) how the agent-turn order is determined. Debate-based systems have been studied in prior work \citep{du2024debate,liang2024divergent,wu2024autogen,li2023camel,swedebate2025,chan2023chateval}; the contribution here is the assessment of these protocols in a controlled mechanism-level comparison in which prompts, decoding settings, and event inputs are matched across protocols, and can highlight their strengths. The analysis focuses on three protocol-sensitive metrics in multi-agent debate: peer-reference rate (PRR; explicit uptake of peer claims), argument diversity (AD; lexical breadth across responses), and consensus formation (CF; reduction in forecast variance across rounds).

The analysis is organized around two hypotheses. \textbf{H1} states that the WR protocol exhibits stronger interaction than the CR protocol, as measured by PRR. \textbf{H2} states that, relative to CR, the RA-CR protocol increases convergence, as measured by CF. These hypotheses define the main claim of the paper: same-round visibility and adaptive participation control may lead to trade-offs between interaction and convergence. Comparisons involving \emph{No-Interaction} are retained as baseline contrasts rather than primary confirmatory tests.

A central component in many multi-agent debate systems is the use of a judge model, by which agent outputs are evaluated, ranked, or scored according to predefined criteria. These evaluations provide structured preference signals that can be used not only for selection but also for learning. In this context, reinforcement learning from AI feedback (RLAIF) can be integrated by treating judge-derived preferences as reward signals to guide agent behavior. This integration is suitable in debate settings, where multiple candidate responses are generated and compared, that allows scalable automated feedback without requiring human annotation. This integration is emerging in multi-agent debate and can be an extension of the protocols used here.

Four main contributions are presented. First, a controlled protocol comparison is provided that studies context availability and agent-turn-order policy under matched prompting and decoding, providing evidence for a trade-off between interaction and convergence. Second, a compact evaluation frame is introduced that separates interaction-sensitive metrics from convergence-sensitive metrics using a small, interpretable metric set. Third, a reproducible orchestration framework with support for reinforcement learning from AI feedback (RLAIF) is presented, which can be transferred to other domains by changing role instructions and event corpora. Fourth, practical deployment guidance is derived: interaction-rich and convergence-oriented protocols should be selected conditionally, motivating complexity-triggered invocation policies rather than always-on debate.

\section{Methods}
\subsection{Dataset}
The dataset is based on the Federal Reserve Economic Data (FRED) series\\
\texttt{CORESTICKM159SFRBATL}, which reports the Sticky Price Consumer Price Index excluding food and energy and is associated with the Federal Reserve Bank of Atlanta \citep{fredcorestick2026}. The raw series provides 121 monthly observations spanning January 2016 through January 2026. In its original form, the series contains dates and monthly inflation values only. To provide additional context for the agents, two additional columns were appended during dataset construction: a \emph{major world event} field and an \emph{inflation relation confirmed} field. Both fields were generated using Google Gemini on March 27, 2026 to provide a salient same-month event and a short note indicating whether a plausible relation to sticky core inflation was present. These annotations are heuristic, are released in the project dataset, and should not be treated as a standard economic benchmark. Two representative examples are given in Supplementary Table~2.

\subsection{Experimental setup}
After constructing the event-annotated dataset, experiments are run under matched prompting and decoding conditions. Fixed base decoding settings (temperature $0.4$), neutral role names (Agent A, Agent B, Agent C), and normalized prompt structure are used across conditions. In implementation terms, these agents are not separate software workers with external tools or memory stores; they are role-conditioned LLM instances whose interaction is mediated entirely through the debate protocol and the orchestrator. Sentence-BERT (SBERT; \texttt{sentence-transformers/all-MiniLM-L6-v2}) is used with greedy max-min selection to construct a diverse top-20 subset from the 121-event dataset. For each event, one experiment is conducted with two debate rounds and five random seeds, yielding 20 $\times$ 5 $= 100$ matched event-by-seed units across protocols. In the main experiment, three different models are used (one per agent), and a separate model is used as the judge: Agent A=\texttt{llama3.2:latest} (architecture: llama; 3.2B; context 131072; quantization \texttt{Q4\_K\_M}), Agent B=\texttt{qwen2.5:3b} (architecture: qwen2; 3.1B; context 32768; quantization \texttt{Q4\_K\_M}), Agent C=\texttt{gpt-oss:20b} (architecture: gptoss; 20.9B; context 131072; quantization \texttt{MXFP4}), and judge=\texttt{mistral:latest} (architecture: llama; 7.2B; context 32768; quantization \texttt{Q4\_K\_M}). These values are reported from the local \texttt{ollama show} metadata for the model tags used in the run. Embedding lengths are omitted here because they were not varied or analyzed in the study.
The code release for the study is available at \url{https://github.com/ramtinz/multi-agent-debate-protocols}.

\begin{figure}[H]
\centering
\includegraphics[width=0.74\linewidth,trim=6 6 6 6,clip]{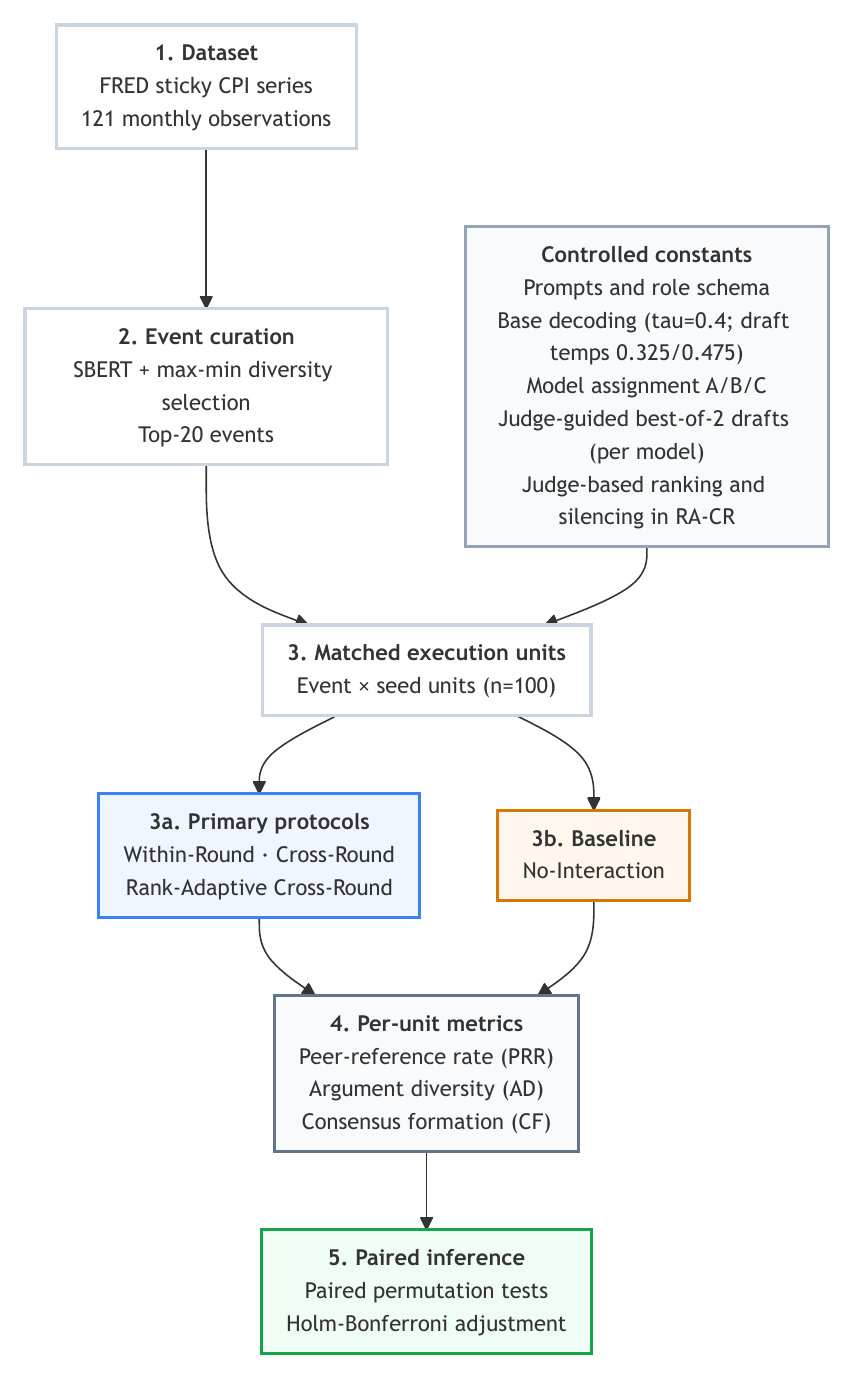}
\caption{Controlled workflow for dataset construction, matched protocol execution, metric computation, and paired inference. Protocol is the manipulated factor, while prompts, decoding settings, model assignment, and judge-guided candidate selection are held fixed across conditions.}
\label{fig:design}
\end{figure}

Figure~\ref{fig:design} summarizes the controlled repeated-measures design. The figure separates manipulated and fixed factors: protocol is the manipulated factor, while prompts, base decoding settings, model assignment, and judge-guided candidate selection are held constant across conditions. Execution is organized as matched event $\times$ seed units ($n=100$), meaning the same 20 events are run with the same five seeds under each protocol, followed by per-unit metric computation and paired inference.

\subsubsection*{Judge usage and response generation policy}
In the primary protocol comparison, each turn uses judge-reranked generation with two candidates. The reranking is \emph{candidate-vs-candidate within the same agent turn}, not agent-vs-agent ranking. Concretely, two drafts are generated from the same prompt with minor temperature perturbations around the same base temperature, both drafts are scored by the judge on a Likert 1--5 rubric, scores are mapped to $[0,1]$, and the top-scored draft is retained. This step therefore selects one of two responses from the same model call context. Small temperature variations are introduced to ensure sufficient candidate diversity; using exactly the same temperature for both would reduce candidate diversity and make reranking less informative. This inference-time filter is used to mitigate stochastic sampling variance and make turn quality more consistent without changing prompt templates, model identities, or event inputs.

One judge model is used in the main experiment (\texttt{mistral:latest}). It has two roles: a condition-uniform intra-turn reranker that selects the better of two candidates for each turn in all protocols, and an adaptive-only controller for next-round ranking and silencing in \emph{RA-CR}. This matters because the adaptive protocol is affected by the judge in an additional way beyond candidate selection. In the upgraded scheduler, rank-adaptive order is quality-biased but randomized each round (rather than deterministic), and the lowest-ranked agent is silenced in the next round under the configured policy. Both WR and CR preserve shuffled agent-turn order to avoid fixed-order artifacts. Judge validity is checked before the main run using five handcrafted event-matched relevant-vs-irrelevant comment pairs; representative examples are given in Supplementary Table~3. For each event, one relevant comment and one irrelevant comment are paired with the same event prompt, ensuring that the comparison depends on the quality of the analysis text rather than on numeric differences. The judge scores both comments independently, and a pair is counted as strictly correct when the relevant comment receives the higher score by at least 0.05. Under this procedure, relevant comments are preferred in 5/5 pairs (strict pairwise accuracy $=1.00$). This check is intended only as a minimal sanity test, not as a comprehensive validation of judge quality. Judge-count sensitivity is additionally examined with the alternate-judge ablation, in which the judge model is replaced with \texttt{llama3.2:latest}. The optional RLAIF components discussed later are supporting infrastructure and are not part of the main workflow shown in Figure~\ref{fig:design}.

\begin{figure}[H]
\centering
\captionsetup[subfigure]{font=small}
\begin{subfigure}[t]{0.495\linewidth}
\centering
\vspace{0pt}
\includegraphics[width=\linewidth,trim=2 2 2 2,clip]{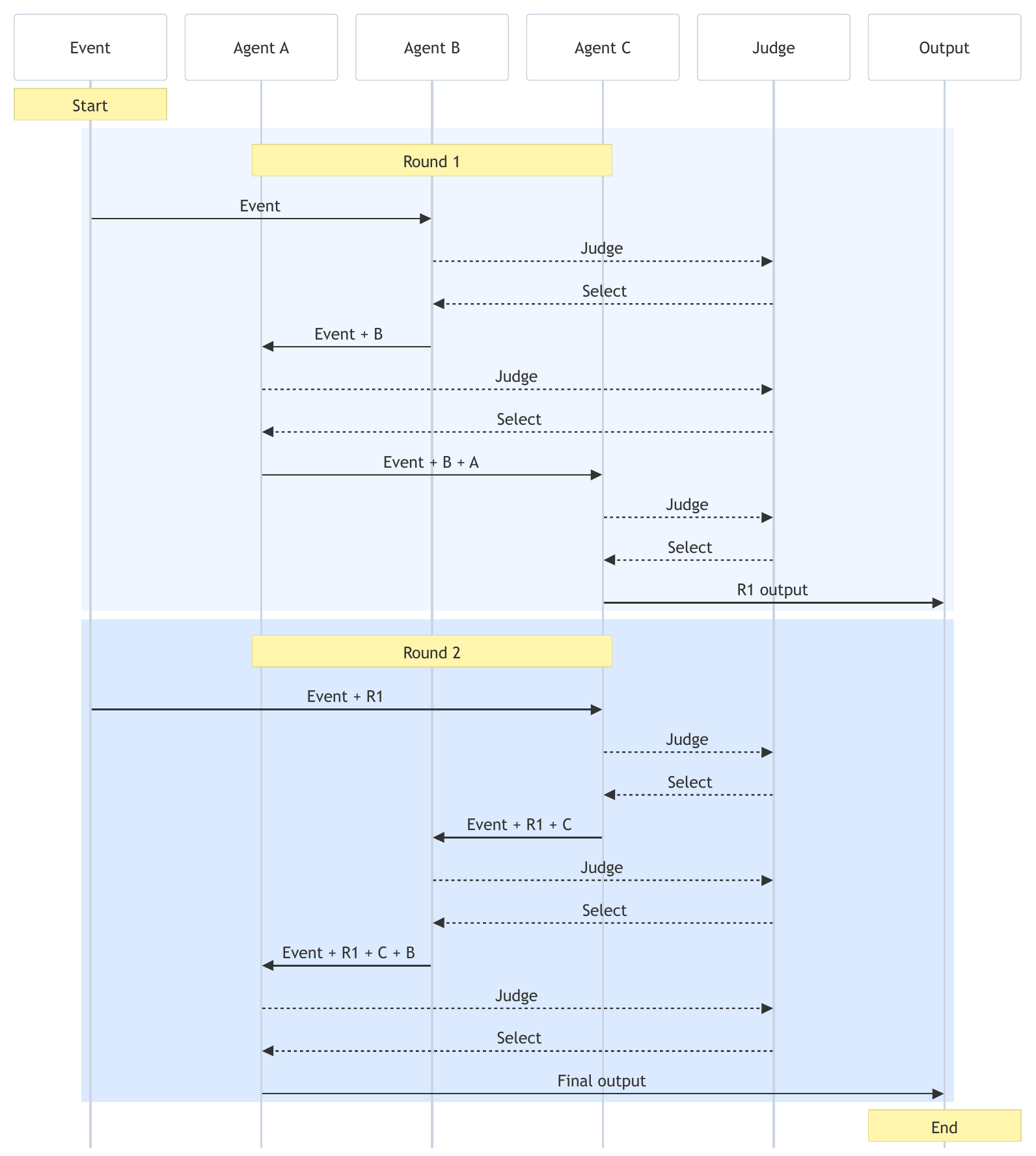}
\caption{Within-Round (WR)}
\label{fig:proto_a}
\end{subfigure}
\hfill
\begin{subfigure}[t]{0.495\linewidth}
\centering
\vspace{0pt}
\includegraphics[width=\linewidth,trim=2 2 2 2,clip]{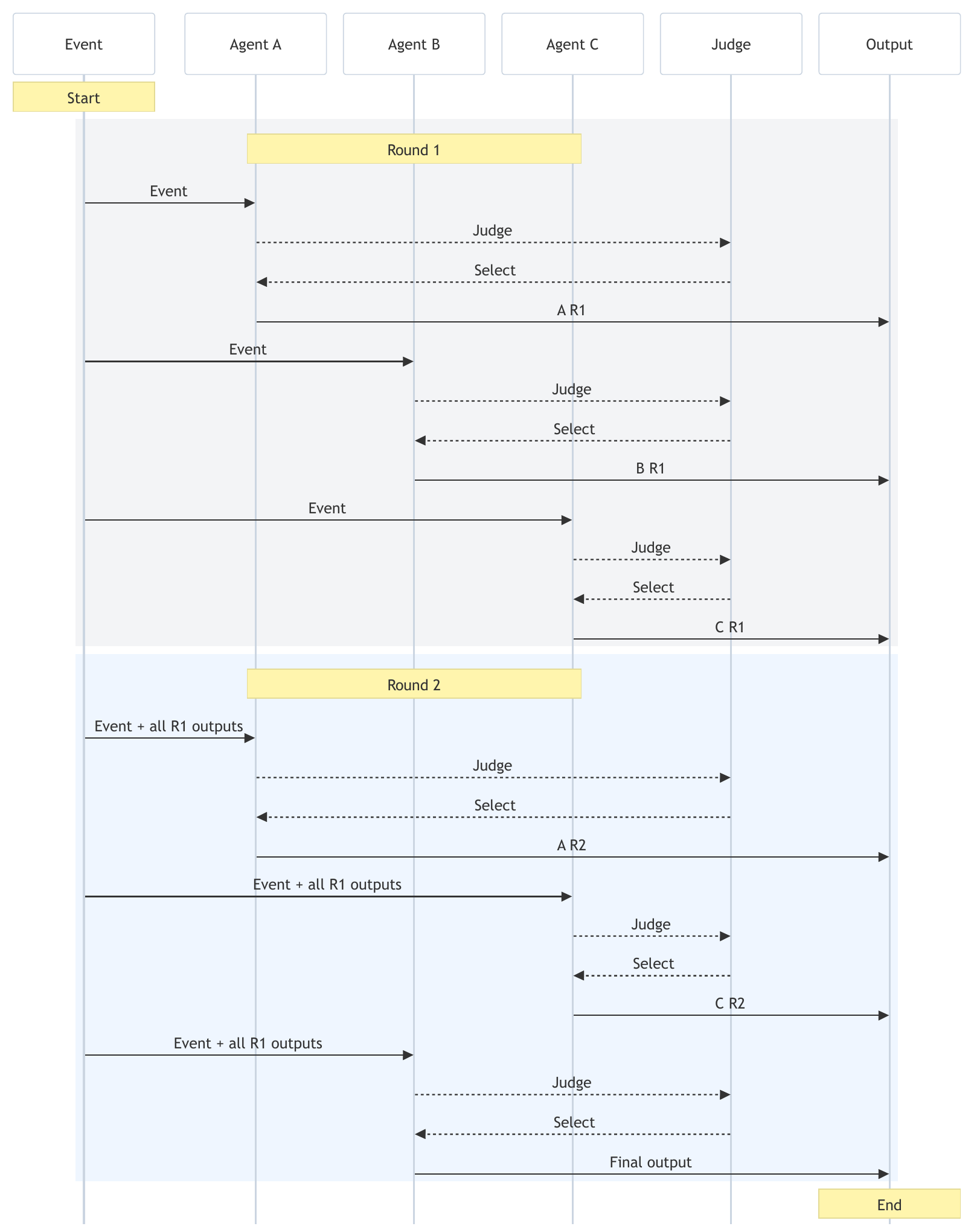}
\caption{Cross-Round (CR)}
\label{fig:proto_b}
\end{subfigure}
\vspace{0.35em}
\begin{subfigure}[t]{0.495\linewidth}
\centering
\vspace{0pt}
\includegraphics[width=\linewidth,trim=2 2 2 2,clip]{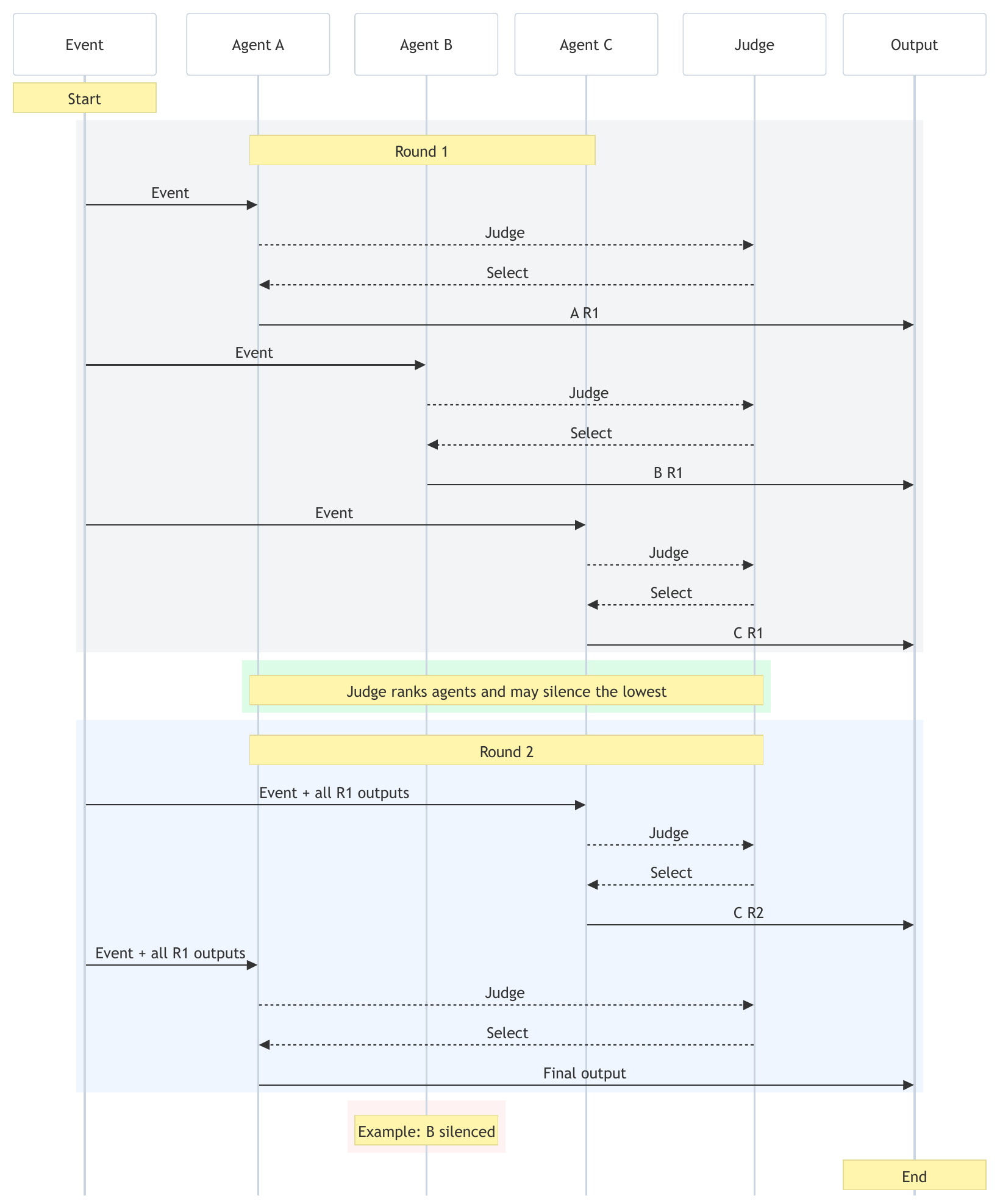}
\caption{Rank-Adaptive Cross-Round (RA-CR)}
\label{fig:proto_c}
\end{subfigure}
\hfill
\begin{subfigure}[t]{0.495\linewidth}
\centering
\vspace{0pt}
\includegraphics[width=\linewidth,trim=2 2 2 2,clip]{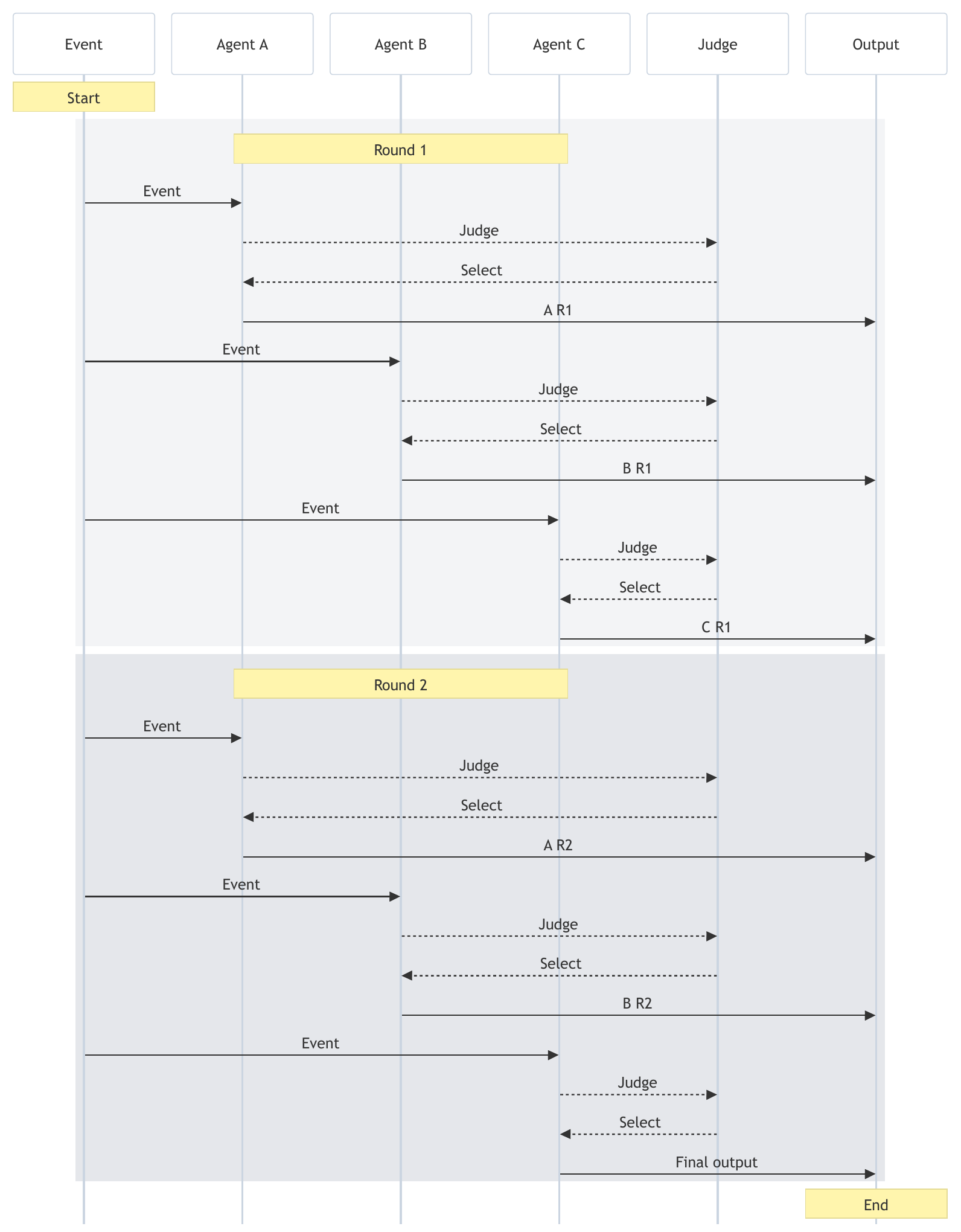}
\caption{No-Interaction (NI)}
\label{fig:proto_d}
\end{subfigure}
\caption{Protocol comparison under a shared two-round timeline for WR, CR, RA-CR, and NI.}
\label{fig:protocols}
\end{figure}

Figure~\ref{fig:protocols} outlines the four debate protocols, with two rounds. Each round has one turn per agent. In panel (a), the full within-round cycle is shown explicitly: event $\rightarrow$ first agent turn $\rightarrow$ second agent turn with event+prior output(s) $\rightarrow$ third agent turn with event+all prior same-round output(s). The key distinction is that WR allows same-round visibility, whereas CR disallows same-round visibility and conditions round-2 turns on round-1 outputs only. RA-CR adds judge-scored, rank-biased randomized ordering and optional silencing on top of CR. NI keeps both rounds self-only. A single output box in each panel collects round outputs and final outputs. Dashed judge paths mark per-turn candidate reranking: candidate drafts are sent to the judge, the selected draft is returned, and that selected draft is emitted as turn output. The green block in panel (c) marks adaptive scheduling and silencing, which is unique to RA-CR. Blue blocks denote peer-visible within-round flow. Agent labels A/B/C identify the three agents; shown orders are examples (shuffled in a, b, d; rank-biased randomized in c). The panels isolate two design dimensions: peer-context horizon and scheduling policy, which jointly determine the interaction--convergence trade-off.

\subsection{Metrics}
Three metrics are presented to evaluate the effects of the debate protocols: peer-reference rate (PRR), argument diversity (AD), and consensus formation (CF). Their definitions are as follows.

\paragraph{Peer-reference rate (PRR).}
A peer reference is counted when another agent is mentioned together with a stance word (e.g., agree or disagree). This rule-based metric is introduced here as an operational measure of explicit cross-agent uptake. Peer names are matched from the peer set (\emph{Agent A}, \emph{Agent B}, \emph{Agent C}); the stance lexicon is fixed to \{\emph{agree}, \emph{disagree}, \emph{challenge}, \emph{support}\}. Argument diversity uses lowercase alphabetic tokens of length $\ge 3$ and Jaccard dissimilarity over pairwise token sets. Numeric forecasts are extracted using simple rules: the number following the \emph{Impact:} line is parsed if present; otherwise, the first signed percentage pattern in the response is parsed. Non-parseable outputs are treated as missing and excluded from variance and shift calculations.
\begin{equation}
\mathrm{PRR} = \frac{1}{N} \sum_{i=1}^{N} \mathbf{1}\{\text{turn } i \text{ contains a valid peer mention and a stance word}\}.
\end{equation}
Here, $N$ is the number of agent turns in the analysis sample, and the indicator equals 1 only when both conditions are satisfied in the same turn. Because the rule is deterministic once the peer set and stance lexicon are fixed, PRR is reproducible for a given transcript set.

\paragraph{Argument diversity (AD).}
Argument diversity is defined as a measure of how different the agents' responses are, based on token overlap. For token sets $\{A_i\}_{i=1}^{n}$, where $A_i$ is the lowercase alphabetic token set extracted from response $i$,
\begin{equation}
\mathrm{AD} = \frac{2}{n(n-1)}\sum_{i<j}\left(1 - J(A_i,A_j)\right),
\end{equation}
where $J(A_i,A_j)=\frac{|A_i\cap A_j|}{|A_i\cup A_j|}$ is Jaccard similarity.
Higher AD means lower lexical overlap across agent arguments (broader exploration), not necessarily better task performance.

\paragraph{Consensus formation (CF).}
Reduction in forecast variance across rounds is used as a measure of consensus formation. The intuition is that if agents are converging on a shared understanding or agreement, their numeric forecasts should become more similar over rounds, leading to reduced variance. CF is defined as the relative reduction in cross-agent variance from the first round to the final round:
\begin{equation}
\mathrm{CF} = \max\!\left(0,\min\!\left(1,1-\frac{\sigma^2_{r=R}}{\sigma^2_{r=1}}\right)\right).
\end{equation}
Here, $\sigma^2_{r=1}$ and $\sigma^2_{r=R}$ denote the cross-agent variance of extracted numeric forecasts in the first and final rounds, respectively. CF increases when disagreement shrinks over rounds. When first-round variance is numerically near zero, a deterministic edge-case rule is used: CF$=1$ if final-round variance is also near zero, else CF$=0$.

\begin{table}[H]
\centering
\caption{Primary metric set used in the main analysis.}
\label{tab:metric_ranges}
\begin{tabular}{lp{2.2cm}p{2.4cm}p{6.0cm}}
\hline
Metric & Operational range & Better direction & Interpretation \\
\midrule
\text{PRR} & $[0,1]$ & Higher & More explicit response to peers \\
\text{AD} & $[0,1]$ & Context-dependent & More lexical variation across responses \\
\text{CF} & $[0,1]$ & Higher & More agreement by the final round \\
\bottomrule
\end{tabular}
\end{table}

\subsection{Statistical analysis}
For each condition mean, a $95\%$ bootstrap confidence interval is reported. Pairwise inference uses two-sided paired permutation tests over matched units. Holm--Bonferroni correction (a step-wise multiple-comparison adjustment) is applied within each protocol comparison family across the three primary metrics, and $p_{\mathrm{HB}}<0.05$ is treated as statistically significant. All primary metrics are bounded in $[0,1]$ and can exhibit non-Gaussian paired differences; accordingly, nonparametric paired permutation tests and bootstrap confidence intervals are used.

\section{Results}

\subsection{Main metric patterns and hypothesis support}
\begin{figure}[H]
    \centering
    \includegraphics[width=0.98\linewidth]{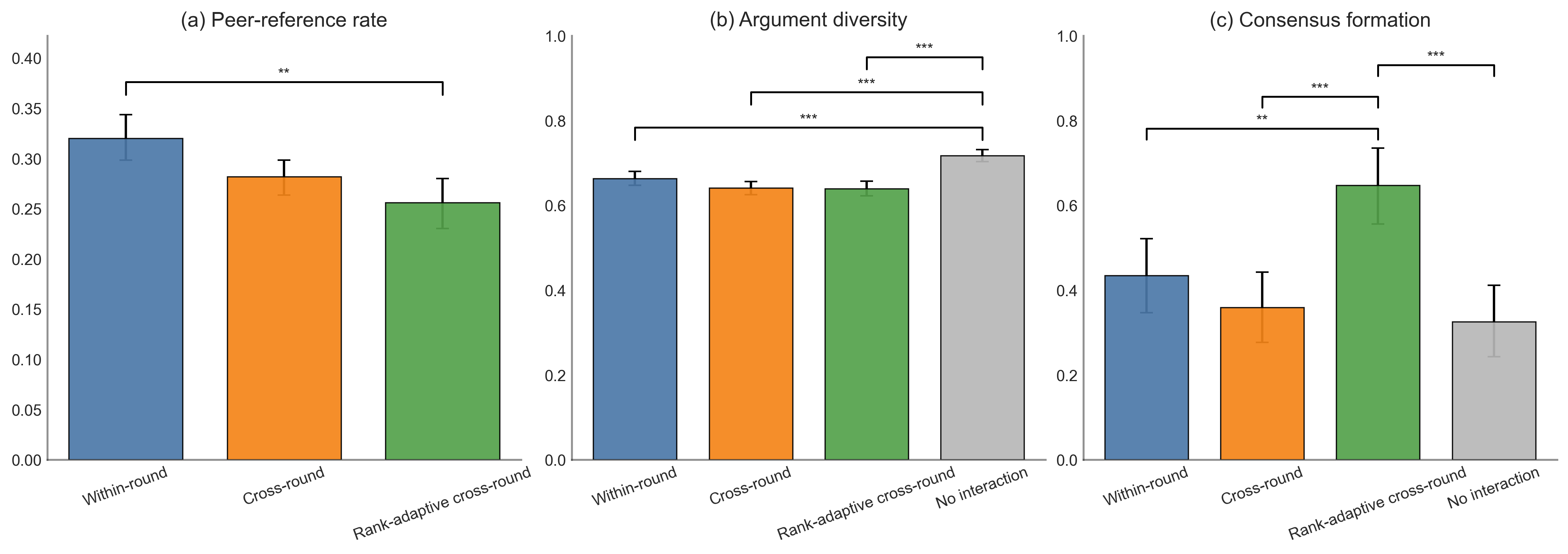}
    \caption{Aggregate comparison of the three primary debate protocols plus the No-Interaction baseline from the SBERT top-20, five-seed aggregate (100 matched units per condition): (a) PRR, (b) AD, and (c) CF. Bars show condition means with 95\% bootstrap confidence intervals. NI is omitted from panel (a) because PRR is structurally zero under no peer visibility. The figure highlights the central trade-off of the study: WR is stronger on interaction-oriented metrics, whereas RA-CR is stronger on convergence-oriented metrics. Significance markers: $*$ $p<0.05$, $**$ $p<0.01$, $***$ $p<0.001$ after Holm--Bonferroni adjustment.}
    \label{fig:aggregate}
\end{figure}

The three primary metrics across the three main protocols, together with the No-Interaction baseline, are presented in Figure~\ref{fig:aggregate}. Average results are reported over $n=100$ matched runs in Figure~\ref{fig:aggregate}. Supplementary Table~1 lists the corresponding condition means and bootstrap confidence intervals. WR has the highest PRR, NI has the highest AD, and RA-CR has the highest CF. Panel (a) shows stronger interaction-oriented behavior for WR, determined by PRR, whereas panel (c) shows that RA-CR is the strongest condition on CF. The paired tests indicate that RA-CR is significantly higher than both WR and CR on CF, and also higher than NI. No significant difference is observed across WR, CR, and RA-CR in AD. In panel (a), NI is omitted because PRR is structurally zero when no peer messages are visible. The higher PRR achieved by WR is consistent with the protocol design: later agents can directly observe and cite earlier same-round outputs, whereas CR conditions defer peer visibility to the next round and RA-CR further reduces some opportunities for explicit peer reference through round-2 silencing.

Pairwise comparisons for the primary protocol metrics are reported in Table~\ref{tab:tests}. The results show that RA-CR has the strongest convergence profile in this experiment: CF is significantly higher for RA-CR than for CR, WR, and NI. A significantly higher PRR is also observed in WR than in RA-CR. For PRR, the WR vs CR and CR vs RA-CR contrasts remain directional but do not cross the Holm--Bonferroni significance threshold ($p_{\mathrm{HB}}=0.0691$ and $0.0510$, respectively). Relative to NI, all three debate protocols show significantly lower AD, and RA-CR also shows significantly higher CF.

\begin{table}[H]
\centering
\caption{Pairwise comparisons for the primary three-metric set after Holm--Bonferroni adjustment. $\Delta$ is the paired mean difference (first minus second). NI contrasts for PRR are omitted because PRR is structurally zero under no peer visibility.}
\label{tab:tests}
\begin{tabular}{llcc}
\toprule
Comparison & Metric & $\Delta$ & $p_{\mathrm{HB}}$ \\
\midrule
WR vs RA-CR & PRR & $+0.064$ & $0.0056$ \\
WR vs RA-CR & AD & $+0.024$ & $0.1045$ \\
WR vs RA-CR & CF & $-0.212$ & $0.0064$ \\
\midrule
WR vs CR & PRR & $+0.038$ & $0.0691$ \\
WR vs CR & AD & $+0.023$ & $0.1596$ \\
WR vs CR & CF & $+0.075$ & $0.7989$ \\
\midrule
CR vs RA-CR & PRR & $+0.026$ & $0.0510$ \\
CR vs RA-CR & AD & $+0.001$ & $1.0000$ \\
CR vs RA-CR & CF & $-0.288$ & $<0.001$ \\
\midrule
WR vs NI & AD & $-0.054$ & $<0.001$ \\
WR vs NI & CF & $+0.109$ & $0.2216$ \\
\midrule
CR vs NI & AD & $-0.077$ & $<0.001$ \\
CR vs NI & CF & $+0.034$ & $1.0000$ \\
\midrule
RA-CR vs NI & AD & $-0.078$ & $<0.001$ \\
RA-CR vs NI & CF & $+0.322$ & $<0.001$ \\
\bottomrule
\end{tabular}
\end{table}

With respect to the stated hypotheses, H2 is supported by the paired tests: RA-CR has the highest CF and exceeds CR on this metric under Holm--Bonferroni adjustment. H1 receives directional but not conclusive support: WR has higher mean PRR and AD than CR, but neither contrast remains significant after Holm--Bonferroni adjustment. Accordingly, the present results do not provide conclusive statistical evidence for the WR-versus-CR contrast.

\section{Discussion}
The findings partially support the hypotheses and highlight the differences the debate protocols can make. Stronger interaction-rich quality signals are produced by WR, whereas stronger convergence-oriented signals are produced by quality-gated RA-CR. These results reinforce the importance of collaborative reasoning and communication topology in multi-agent debate. When consensus formation is the primary system objective, RA-CR is the most promising protocol among those tested. Protocol choice materially changes system behavior, and RA-CR provides a clear advantage when convergence is prioritized under controlled conditions.

An actionable taxonomy follows directly from the results: protocols can be located along two axes, \emph{interaction richness} (peer uptake and cross-agent engagement) and \emph{convergence pressure} (variance reduction and consensus formation). WR occupies the interaction-rich region, while RA-CR shifts toward convergence-oriented coordination. This framing implies that benchmark conclusions can depend on protocol choice and that the debate protocol should therefore be reported and tuned as an explicit experimental variable rather than treated as a fixed background implementation detail.

Existing work has studied the utility of debate and multi-agent collaboration \citep{du2024debate,liang2024divergent,wu2024autogen,li2023camel,hong2023metagpt}. More recent systems have explored dynamic team formation and communication structures for LLM-based agents \citep{liu2024dylan}, competitive debate in software engineering \citep{swedebate2025}, debate-style evaluation \citep{chan2023chateval}, automated protocol and topology optimization over agentic workflow spaces \citep{maas2025,zhang2024aflow}, and adaptive self-improving agents \citep{shinn2023reflexion}. Additional strands include equilibrium-based coordination \citep{xie2025equilibrium}, selective debate triggering for token efficiency \citep{fan2026imad}, and focused analyses of debate-versus-vote decision rules \citep{choi2025debatevote}.

Among these, ChatEval is especially close to this study because it treats each LLM as an agent and compares alternative debate communication strategies \citep{chan2023chateval}. Its \emph{One-By-One} setting is broadly analogous to a within-round protocol with sequential same-round visibility, whereas its \emph{Simultaneous-Talk} setting is closer to a cross-round design in which same-round peer messages are not visible during generation. ChatEval, however, studies these choices as components of an evaluator architecture together with diverse role prompts and an optional summarizer, and its final outputs are aggregated by annotator voting or averaging rather than by directly measuring convergence. By contrast, the present study isolates the debate protocol itself under matched prompts, decoding, and model assignment, and introduces RA-CR as an adaptive extension in which judge-based ranking changes later-round participation and turn order. In that sense, RA-CR can be understood as a controller-level extension that could in principle be layered on top of ChatEval-style debate systems, although that transfer would still require task-specific validation.

Related work has also examined post hoc attribution in multi-agent debate. IntrospecLOO \citep{cui2025introspecloo} approximates leave-one-out agent contributions by adding an extra querying round after the debate, in which agents update their answers while ignoring a designated peer. This reduces the cost of full leave-one-out evaluation, but it does not modify the debate protocol during the interaction itself. RA-CR differs in both purpose and timing: it changes the protocol in situ by using a judge to rank agents, reorder later-round participation, and silence the lowest-ranked agent. IntrospecLOO is therefore better understood as a complementary attribution method than as a direct alternative to protocol-level adaptation.

In parallel, contemporary agent frameworks often use reusable engineering patterns such as explicit planning tools, subagent delegation, persistent memory, and context offloading \citep{langchainmultiagent2026}. Recent surveys also emphasize that LLM-based agent evaluation spans a broad spectrum of architectures, from relatively lightweight orchestrated role-play settings to more tool-using and environment-interacting systems \citep{yehudai2025survey}. Recent domain- and mechanism-specific extensions further motivate this focus, including claim-verification debate with explicit moderator design \citep{he2026debatecv}, asymmetric peer-prediction protocols intended to break correlated-error regimes \citep{liu2026acemad}, dynamic role assignment for capability-aware debate orchestration \citep{zhang2026dynamic}, context-learning approaches that adapt discussion context across rounds \citep{hua2026context}, evaluation-oriented debate frameworks with explicitly defined communication strategies \citep{chan2023chateval}, and explicit thinking-mode decomposition for interpretable collaboration \citep{hefeng2026dimo}. Broader surveys and comparative audits also emphasize that cost, topology, and bias remain unresolved design constraints in practical multi-agent debate pipelines \citep{smit2024mad,tillmann2025review}.

Broader controller-style methods, including adaptive routing, dynamic role assignment, and context-conditioned coordination, were considered during the development of this study \citep{yehudai2025survey,liu2024dylan,zhang2026dynamic,hua2026context}. They were not treated as part of the main analysis because the present paper is intentionally scoped to a controlled comparison of the primary debate protocols under matched conditions.

\subsection{Limitations}

There are some limitations to consider regarding the scope of this study, and potential usage and extensions of the findings. First, the experiments are conducted in one macroeconomic setting with a top-20 event subset; external validity across other domains and larger model classes therefore remains to be tested. Second, the judge has a dual role as both reranker and adaptive controller, which means some protocol effects may be partly judge-mediated; the ablations reduce this concern but do not remove it entirely. Third, the analysis focuses on model-based and protocol-based metrics rather than human evaluation, thus the reported quality gains should be interpreted within that measurement frame.

\subsection{Future work}
The most immediate extensions are broader judge validation, human evaluation of final outputs, richer semantic diversity metrics, and replication across larger model families and other reasoning domains. A second direction is to study protocol selection more directly, for example by learning when interaction-rich protocols justify their additional cost and when simpler protocols are sufficient. A third direction is to connect debate protocols more directly to RLAIF and scalable oversight, for example by studying when multi-agent deliberation provides information that is not recovered by single-model feedback optimization alone, and by training deliberative policies with judge-derived rewards in negotiation-style settings \citep{young2026knowledge,anantaprayoon2026learning}.

\section{Conclusion}
A controlled comparison of multi-agent debate protocols for macroeconomic event analysis is presented, using a semantically diverse top-20 subset and five seeds. Under matched prompts and decoding, protocol choice changes behavior in systematic ways: WR yields the highest peer-reference rate, whereas RA-CR yields the highest consensus formation. Pairwise tests show that RA-CR exceeds WR, CR, and NI on CF, while WR exceeds RA-CR on PRR. These results identify a clear trade-off between interaction and convergence and indicate that RA-CR is the most promising protocol among those tested when consensus formation is prioritized. More broadly, the debate protocol should be treated and reported as an explicit design variable in MAD evaluation.

\appendix

\section{Appendix: Pseudocode and Supplementary Material}
\setcounter{algorithm}{0}
\renewcommand{\thealgorithm}{\Roman{algorithm}}
\setcounter{table}{0}
\renewcommand{\thetable}{\arabic{table}}
\captionsetup[table]{labelformat=empty,justification=raggedright,singlelinecheck=false}

\subsection{AI-judge best-of-$N$ selection}
\begin{algorithm}[H]
\caption{AI-judge best-of-$N$ response selection}
\label{alg:bestofn}
\begin{algorithmic}[1]
\Require Prompt $p$, generation model $\pi$, judge model $J$, candidates $N$, base temperature $\tau$
\For{$i = 1$ to $N$}
    \State $t_i \leftarrow \tau + (i - (N-1)/2) \times 0.15$ \Comment{temperature jitter}
    \State $r_i \leftarrow \pi(p; t_i)$ \Comment{sample candidate response}
    \State $s_i \leftarrow J(p, r_i)$ \Comment{judge scores on Likert 1--5 $\to$ $[0,1]$}
\EndFor
\State \Return $r^* = \arg\max_i s_i$ \Comment{select highest-scored candidate}
\end{algorithmic}
\end{algorithm}

\subsection{Rank-adaptive cross-round scheduling}
\begin{algorithm}[H]
\caption{Rank-adaptive cross-round scheduling}
\label{alg:rankadaptive}
\begin{algorithmic}[1]
\Require Agent set $\mathcal{A}$, round count $R$
\For{$r = 1$ to $R$}
    \If{$r = 1$}
        \State Randomly permute $\mathcal{A}$ to obtain agent order
    \Else
        \State Score round $r{-}1$ responses with a local LLM judge (\texttt{mistral}, Likert 1--5 rubric $\to$ $[0,1]$)
        \State Sample next-round order with quality-biased randomization (higher score $\Rightarrow$ higher probability of earlier turn)
        \State Preserve stochastic ordering to avoid deterministic fixed-order artifacts
    \EndIf
    \State Execute debate turns for round $r$
    \State Record per-agent judge scores for next-round ranking
\EndFor
\end{algorithmic}
\end{algorithm}

\subsection{Supplementary table: confidence intervals}
\begin{table}[H]
\caption{Supplementary Table 1. Condition means with bootstrap $95\%$ confidence intervals from the SBERT top-20, five-seed main aggregate ($n=100$ per condition).}
\label{tab:supp_ci}
\begin{tabular}{lccc}
\hline
Condition & Metric & Mean & 95\% CI \\
\midrule
Within-Round & Peer Reference Rate & 0.320 & [0.298, 0.343] \\
Cross-Round & Peer Reference Rate & 0.282 & [0.263, 0.298] \\
Rank-Adaptive Cross-Round & Peer Reference Rate & 0.256 & [0.230, 0.280] \\
Within-Round & Argument Diversity & 0.663 & [0.646, 0.679] \\
Cross-Round & Argument Diversity & 0.640 & [0.625, 0.656] \\
Rank-Adaptive Cross-Round & Argument Diversity & 0.639 & [0.622, 0.656] \\
No-Interaction & Argument Diversity & 0.717 & [0.696, 0.736] \\
Within-Round & Consensus Formation & 0.434 & [0.346, 0.521] \\
Cross-Round & Consensus Formation & 0.359 & [0.276, 0.443] \\
Rank-Adaptive Cross-Round & Consensus Formation & 0.647 & [0.555, 0.734] \\
No-Interaction & Consensus Formation & 0.325 & [0.250, 0.411] \\
\bottomrule
\end{tabular}
\end{table}

\subsection{Supplementary examples: dataset annotations and judge validation}

\begin{table}[H]
\caption{Supplementary Table 2. Example rows from the event-annotated dataset.}
\label{tab:event_examples}
\begin{tabular}{p{1.6cm}p{1.3cm}p{6.6cm}p{5.0cm}}
\hline
Date & Inflation (\%) & Major world event & Inflation relation confirmed? \\
\midrule
2016-02 & 2.54 & The World Health Organization declares the Zika virus outbreak a Public Health Emergency of International Concern. & No confirmed correlation with US sticky price movements. \\
2016-05 & 2.57 & A massive wildfire in Fort McMurray, Alberta, forces the largest evacuation in Canadian history and halts oil sands production. & Minimal; energy is excluded from this index, and the production halt was temporary. \\
\bottomrule
\end{tabular}
\end{table}

Table~\ref{tab:judge_examples} summarizes representative judge-validation items. Each item contains one event prompt paired with one relevant comment and one clearly irrelevant comment; both comments are scored independently by the judge under the same fixed \emph{Impact: +0.2\%} line. The full validation set contains five such pairs.

\begin{table}[H]
\caption{Supplementary Table 3. Representative judge-validation examples used for the pre-run sanity check.}
\label{tab:judge_examples}
\begin{tabular}{p{4.4cm}p{5.0cm}p{5.0cm}}
\hline
Event & Relevant comment & Irrelevant comment \\
\midrule
Fed raises policy rate by 75 bps amid persistent core inflation & The rate hike should cool demand-sensitive sectors over 6--12 months, but sticky services inflation may remain elevated due to wage persistence. & I like hiking and the weather is nice this week. This event reminds me of my vacation photos. \\
Major shipping disruptions in the Red Sea increase global freight costs & Higher freight costs can pass through to goods prices and import channels, adding near-term upside pressure to US sticky core inflation. & My favorite movie soundtrack is underrated and has great violin solos. \\
\bottomrule
\end{tabular}
\end{table}

\bibliographystyle{icml2026}
\bibliography{references}

\end{document}